\documentclass[a4paper,11pt]{article}
\usepackage[utf8]{inputenc}
\usepackage{graphicx}% Include figure files
\usepackage{amsmath}
\usepackage{amssymb}
\usepackage{latexsym}
\usepackage{fullpage}

\begin{document}

\title{The first-digit frequencies in data of turbulent flows}
\author{Damien BIAU \\DynFluid Laboratory, Arts et M\'etiers ParisTech,\\ 151 Boulevard de l'H\^opital,\\ 75013 Paris, France}
\date{damien.biau@ensam.eu}
\maketitle

%%%% Abstract text to be placed here %%%%%%%%%%%%
\begin{abstract}
Considering the first significant digits (noted d) in data sets of dissipation for turbulent flows, 
the probability to find a given number (d=1 or 2 or... 9) would be 1/9 for an uniform distribution. 
Instead the probability closely follows Newcomb-Benford's law, namely P(d)=log(1+1/d). 
The discrepancies between Newcomb-Benford's law and first-digits frequencies in turbulent data are analysed through Shannon's entropy.
The data sets are obtained with direct numerical simulations for two types of fluid flow: an isotropic case initialized with a Taylor-Green vortex and a channel flow.
Results are in agreement with Newcomb-Benford's law in nearly homogeneous cases and the discrepancies are related to intermittent events.
Thus the scale invariance for the first significant digits, which supports Newcomb-Benford's law, seems to be related to an equilibrium turbulent state, 
namely with a significant inertial range.
A matlab/octave program is provided in appendix in such that part of the presented results can easily be replicated.
\end{abstract}
%%%%%%%%%%%%%%%%%%%%%%%%%%%

%--------------------------------------------------------------------------------------------------
\section{Introduction}
%--------------------------------------------------------------------------------------------------

In 1881 Newcomb \cite{Newcomb} observed a rather strange fact: tables of logarithms in libraries tend to be quite dirty at the beginning and progressively cleaner throughout. This seemed indicate that people had more occasion to calculate with numbers beginning with 1 than with other digits. Newcomb concluded that the frequency of the leftmost, nonzero digit $d$ closely follows the probability law:
\begin{equation}\label{NBlaw}
P(d) = \log_{10} \left( 1+ \frac{1}{d} \right) \quad d=1,2,...~ 9
\end{equation}
Hence the numbers in typical statistics should have a first digit of 1 about $30\%$ of the time, but a first digit of 9 only about $4.6\%$ of the time 
(the other values can be found in table \ref{NumValues}). Also, the probability that the first significant digit is an odd number is $60\%$.
This formula is also valid for the digits beyond the first, for example the distribution of the $k$ first digits is given by equation \ref{NBlaw} with 
$d=10^{k-1},...~ 10^{k}-1$. 

Newcomb's article went unnoticed until 1938 when Benford \cite{Benford} has independently followed the same path: starting from observations about logarithm books, he deduced the same law. Benford also noted that this law is base-invariant, the definition (\ref{NBlaw}) is given here with a decimal representation (base-10) for convenience.
Until today this empirical law has been tested against various datasets  ranging from  mathematical curiosity to natural sciences \cite{Hill98,Sambridge}.

Then there have been many attempts to rationalize this empirical law \cite{Raimi,Hill95,Pietronero,Pinkham,Pocheau,Fewster}. 
Recently Fewster \cite{Fewster} has provided an intuitive explanation based on the fact that any distribution has tendency to satisfy Newcomb-Benford's law, as long as the distribution spans several orders of magnitude and as long as the distribution is reasonably smooth.

Pietronero \emph{et al} \cite{Pietronero} have shown that Newcomb-Benford's law is equivalent to the scale invariance of data set.
%The scale-invariance can be stated as follows (see Pietronero \emph{et al} \cite{Pietronero}). 
Multiplying a data set $X$ by a factor $\lambda$ rescales the first-digits distribution such that: 
$ P(d[\lambda X]) = \lambda^m P(d[X]),$ 
where $d[X]$ denotes the first significant digit of the variable $X$ and $m$ is an exponent to be determined.
The general solution of this equation takes the form of power-law.%, $Kx^{-\alpha}$, $\alpha>0$.
The probability distribution $P(d)$, namely the sub-interval of $[1,~ 10)$ occupied by the first digit $d$, is obtained by integration:
$$P(d) = K \int_d^{d+1}~ z^{-\alpha}~ dz $$
\noindent where K is a necessary constant to enforce the condition $\sum P=1$.
Newcomb-Benford's law is recovered for $\alpha=1$, while $\alpha \ne 1$ leads to the generalized distribution found by 
Pietronero \emph{et al.} \cite{Pietronero}:
\begin{equation}\label{generalized}
P(d)= \frac{ (d+1)^{1-\alpha}-d^{1-\alpha} }{10^{1-\alpha} - 1}
\end{equation}
The generalized Newcomb-Benford law has been applied to the distribution of leading digits in the prime number sequence by Luque and Lacasa \cite{Luque}, 
the results have shown an asymptotic evolution toward the uniform distribution ($\alpha=0$).
Note that as particular case $\alpha=1$, Newcomb-Benford's law satisfies a strong invariance, \emph{i.e.} $P(d[\lambda X])=P(d[X])$. 
This result was first demonstrated by Hill \cite{Hill95}. 
As illustration, we consider $X$ such that $d[X]=7$ and $\lambda=2$ as scaling factor, thus $d[\lambda X]=1$, but the secondary digit can only be 4 or 5. 
With the relation (\ref{NBlaw}), it is easy to check the scale-invariance for this example: $P(7)=P(14) + P(15)$.

%%  turbulence
Self-similarity is also a well explored topic in turbulent flows.
The first step was made by Richardson in the late 19th century with the assumption of an universal cascade process of the energy:
the energy input at large scale is successively transferred to finer eddies.
This idea has been refined by Kolmogorov and Obukov, the cascade is then assumed to occur in a space-filling, self-similar way.
Formally there should be an unique scaling exponent for the structure functions $S(r)$ such that $S(\lambda r) = \lambda^h S(r)$.
Today the departures from the Kolmogorov scaling prediction are identified to different causes: the Reynolds number is not large enough, 
the flow is not exactly isotropic or the self-similarity assumption is not valid.
The last point is related to the intermittent feature of the cascade process, resulting in anomalous scaling, or no unique scaling exponent. 
Further informations can be found in the book by Frisch \cite{Frisch}.

%% phrase d'accroche
In the present work, the distribution of first significant digits is used as an alternative statistical tool for analysing turbulent flows.
Newcomb-Benford's law is compared to data sets coming from numerical simulations of the Taylor-Green vortices (homogeneous case) and the plane Poiseuille flow (inhomogeneous case). In the next section the numerical method is presented and the Shannon entropy is introduced as a diagnostic tool. 
Results are presented in the following section. The validity and the possible extensions of the method are discussed in the last section.

%--------------------------------------------------------------------------------------------------
\section{Method}
%--------------------------------------------------------------------------------------------------

The momentum and mass conservation equations applied to the fluid leads to the Navier-Stokes equations, written in non dimensional form:

\begin{equation}\label{eq_NS}
\begin{array}{l}\displaystyle 
\nabla \cdot \textbf{u} = 0, \\[8pt]
%\frac{\partial \textbf{u}}{\partial t} = \textbf{u} \wedge \omega - \nabla \left(p+\frac{1}{2}u^2\right) + Re^{-1} \nabla^2 \textbf{u}
\partial_t \textbf{u} + \textbf{u} \cdot \nabla \textbf{u} = - \nabla p + Re^{-1} \nabla^2 \textbf{u}
\end{array}
\end{equation}
\noindent the flow is assumed incompressible and the Reynolds number ($Re$) is the only control parameter.
% $\mathbf{u}$, $p$ is the pressure

%%  NON-LINEAR
It should be noted that the quadratic nonlinear term in the Navier-Stokes equations can be related to Newcomb-Benford's law.
If we assume a scale invariant data set, then the peculiar Newcomb-Benford's law could be rooted in this non-linear term. 
Let consider the quadratic transformation, $X\rightarrow X^2$,  mimicking the non linear convective term in Navier-Stokes equations \ref{eq_NS}. 
The probability to find the first digit of $X$ in the interval $[d,~ d+1)$ equals the probability to find the first digit of $X^2$ in $[d^2,~ (d+1)^2)$:
$$ \int_{d^2}^{(d+1)^2} z^{-\tilde{\alpha}}~dz = \int_d^{d+1} z^{-\alpha}~ dz $$
leading to $\tilde{\alpha}=(\alpha+1)/2$. For any distribution, characterized by an exponent $\alpha$, the quadratic transformation generates a distribution with an exponent $\tilde{\alpha}$ which is closer to 1 than initial $\alpha$. As consequence the quadratic  non-linear term in Navier-Stokes equations \ref{eq_NS}
acts as a feedback which force any scale invariant distribution (any $\alpha$) toward Newcomb-Benford's law ($\alpha = 1$).
However the underlying reason why scale invariance occurs is still elusive.
As stated by Fewster \cite{Fewster}, currently any argument can explain why a universal law of nature should arise in the first place.

Now we have to choose the observable which will be used to apply the statistical analysis.
The viscous dissipation has been selected:
\begin{equation}\label{diss}
\epsilon = \frac{1}{Re} ~ \frac{1}{2}\left( \frac{\partial u_i}{\partial x_k} + \frac{\partial u_k}{\partial x_i} \right)^2 
\end{equation}

In a practical point of view, the dissipation in internal or external flow fields is directly related to friction factor and drag coefficient respectively.
In addition, for isothermal flows, the dissipation is the only term responsible for entropy production.
Hence the dissipation can be considered as cornerstone for turbulent statistics, for a recent overview see Vassilicos \cite{Vassilicos}. 
The dissipation appears as a natural choice in order to exemplify the first-digit statistics in turbulent flows.

%%%%   ENTROPY
In order to investigate the possible conformance of turbulent data to Newcomb-Benford's law, it is necessary to quantify their discrepancy. 
This can be achieved through different ways. First the usual statistical tools used to test the goodness of fit, for example the $\chi^2$ or chi-square test.
Secondly the search of the optimal exponent $\alpha$ in the generalized distribution, equation (\ref{generalized}), see Luque and Lacasa \cite{Luque}.
Eventually a third choice was retained, stemming from the information theory. 
In 1948 Shannon \cite{Shannon} introduced entropy as a measure of the uncertainty in a random variable:
\begin{equation}\label{entropy_def}
H = -\sum_{d=1}^9 P(d)~ \log_{10} P(d)
\end{equation}

This definition of entropy is associated to lack of information or to uncertainties. A message, in our case a chain of first digits d, occurring with probability $P(d)$, contains the information $I=-\log_{10}P(d)$. Thus Shannon entropy $H$ may be seen as the information averaged over the message space.
That formula can be heuristically explained considering the fact that total information of two independent events (with respective probabilities $P_1$ and $P_2$) is the sum of the two individual information, namely $ I(P_1,~ P_2) = I(P_1)+I(P_2)$, hence the logarithm.
Moreover information is a positive quantity, hence minus sign, and the information contained in certain events vanishes: $I(1)=0$.

\begin{figure}[!ht]
 \centerline{\includegraphics[width=0.5\textwidth]{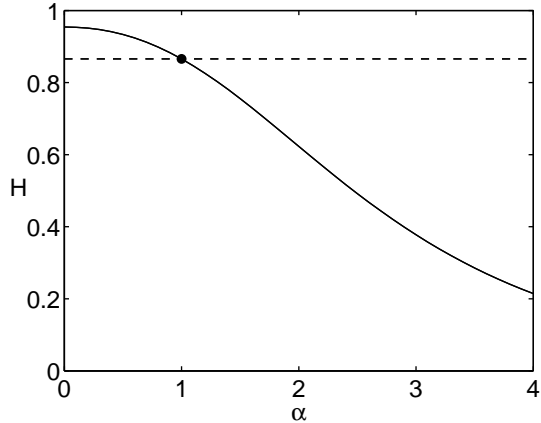}}
 \caption{\label{H_alpha} Entropy for various distributions. Newcomb-Benford's case, marked with a solid circle, corresponds to $\alpha=1$ and $H=0.8657$.}
\end{figure}

The Shannon entropy applied to the general probability distribution (equation \ref{generalized}) is depicted in figure \ref{H_alpha}.

The maximum entropy is achieved when all digits are equally probable.
The principle of maximum entropy leads to the notion of equipartition: the equal spread of probabilities over all possible states of a system. 
In figure \ref{H_alpha}, the maximum value is $H=\log_{10}(9)\approx 0.9542$. Conversely the minimum entropy occurs when one digit is certain, then entropy vanishes. 

%%  EXTENSIVITY
The extensive property of entropy is straightforward if we consider Newcomb-Benford's law for digits beyond the first. 
Hill \cite{Hill98} have shown that the probability $P$ that d (d = 0, 1, ..., 9) is encountered as the n-th ($n>0$) digit is:
$$ P_n(d) = \sum_{ k = 10^{ n - 1 } }^{ 10^{ n } - 1 } \log_{ 10 } \left( 1 + \frac{ 1 }{ 10k + d } \right ). $$
The distributions of the first and second digits are not independent, however, as n increases, 
the distributions of successive digits become independent and rapidly converge to an even distribution (p=0.1 for each of the ten digits).
Thus, for sufficiently large n, the entropy satisfies the extensive property:
$$ H( P_n, ~ P_{n+1} )=H(P_n)+H(P_{n+1}) \rightarrow 2 \quad \mathrm{as} ~ n\rightarrow \infty$$

%--------------------------------------------------------------------------------------------------
\section{Results}
%--------------------------------------------------------------------------------------------------

In the following, two turbulent flows are considered.
First a three-dimensional decaying and homogeneous turbulent flow, initialized with a Taylor-Green vortex.
Then a developed wall-bounded turbulent flow in a plane channel.

%-------------------------------------------------------------------
\subsection{Homogeneous and isotropic case: the Taylor-Green vortex}

We consider a spatially periodic flow, the problem is studied by direct numerical simulation with $256^3$ Fourier modes and the Reynolds number is $Re=1600$.
As initial condition we impose the single-mode Taylor-Green vortex \cite{TGV,Brachet}:
\begin{equation}\label{eq_NS}
\begin{array}{rcl}\displaystyle
u(x,y,z,t=0) & = &  2/\sqrt{3}~ \sin\left( 2\pi/3 \right) ~\sin x~ \cos y~ \cos z \\
v(x,y,z,t=0) & = & -2/\sqrt{3}~ \sin\left( 2\pi/3 \right) ~\cos x~ \sin y~ \cos z \\
w(x,y,z,t=0) & = & 0
\end{array}
\end{equation}

This initial solution generates a very complex flow and evolves into turbulent flow for sufficiently high Reynolds number.
The small scales are generated by three-dimensional vortex stretching, as described in the original article by Taylor and Green \cite{TGV}.
Because of deterministic nature of the Navier-Stokes equations, the flow issuing from a Taylor-Green vortex presents microscopic structure which can be reproduced repeatedly. The results presented hereafter can thus be easily reproduced with the program provided in appendix. %\ref{}

\begin{figure}[!ht]
 \includegraphics[width=0.5\textwidth]{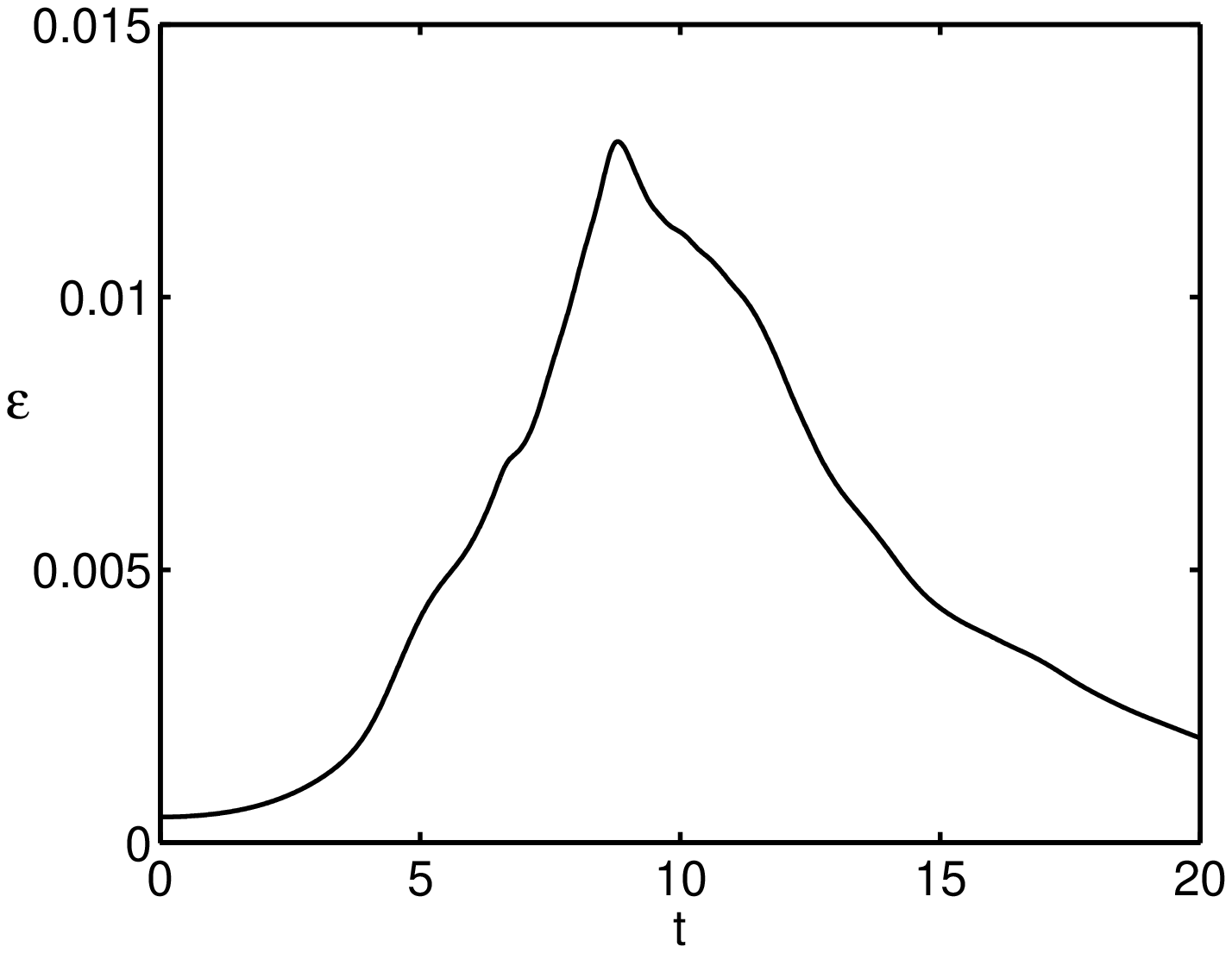}%
 \includegraphics[width=0.5\textwidth]{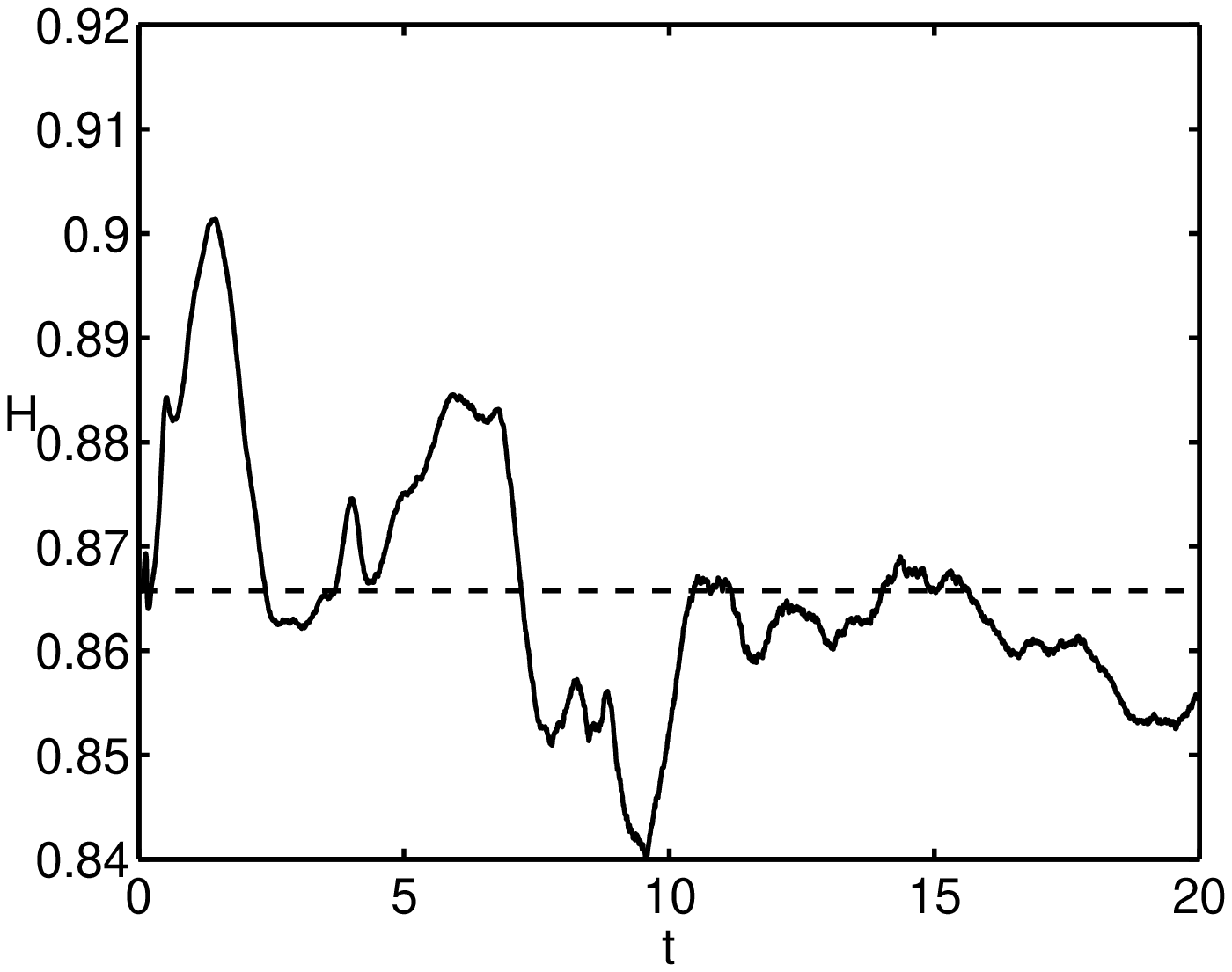}%
 \caption{\label{TGV_H_eps} Left: Rate of energy dissipation $\epsilon$ as a function of time for $Re=1600$.
Rigth: Shannon entropy \emph{vs} t for the Taylor-Green vortex. 
The dotted line stands for the Shannon entropy corresponding to Newcomb-Benford's law, $H=0.8657$.}
\end{figure}

The time evolution of the total dissipation and the conformance of turbulent statistics with Benford's law are presented in figure \ref{TGV_H_eps}.

\begin{figure}[!ht]
 \includegraphics[width=0.5\textwidth]{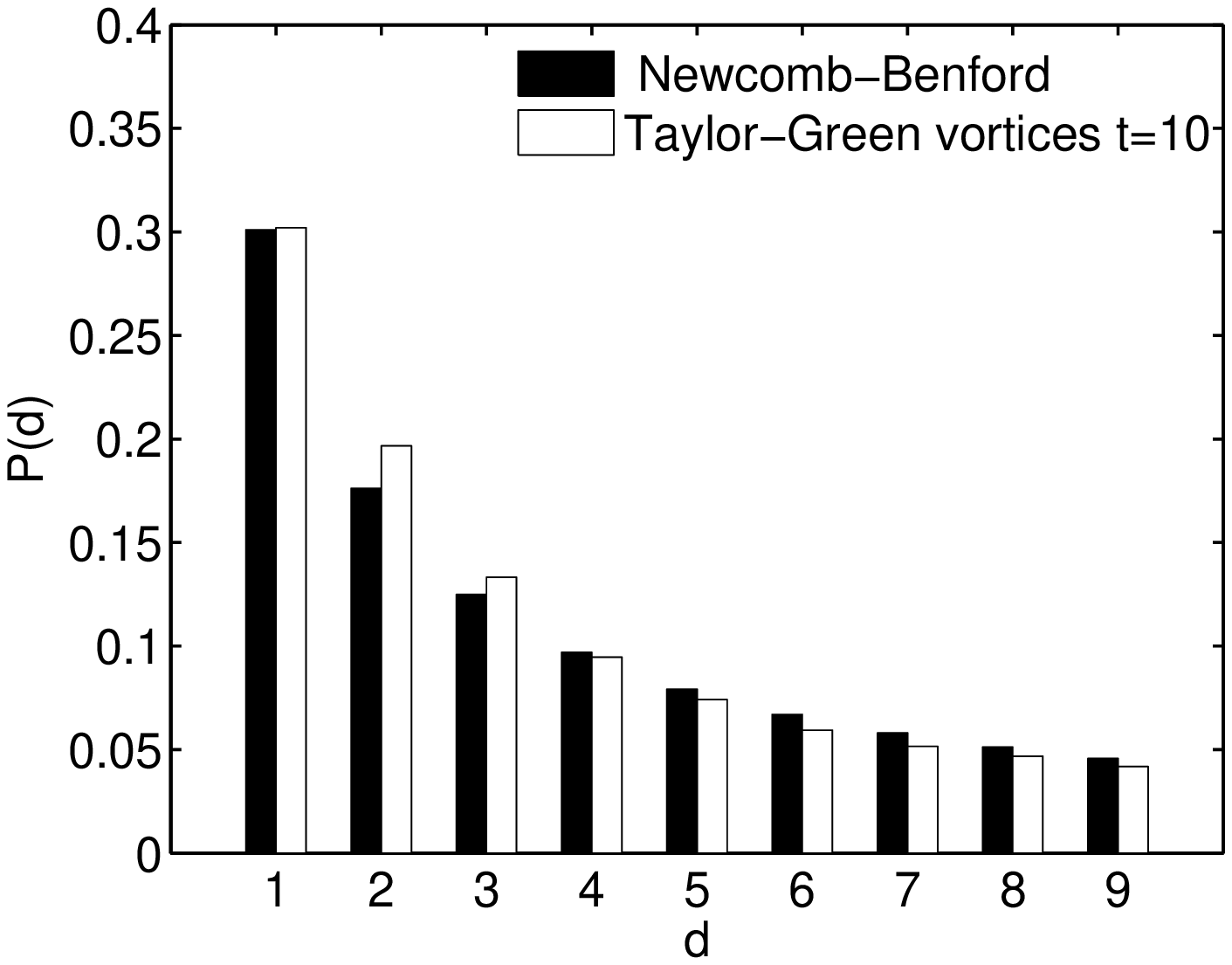}%
 \includegraphics[width=0.5\textwidth]{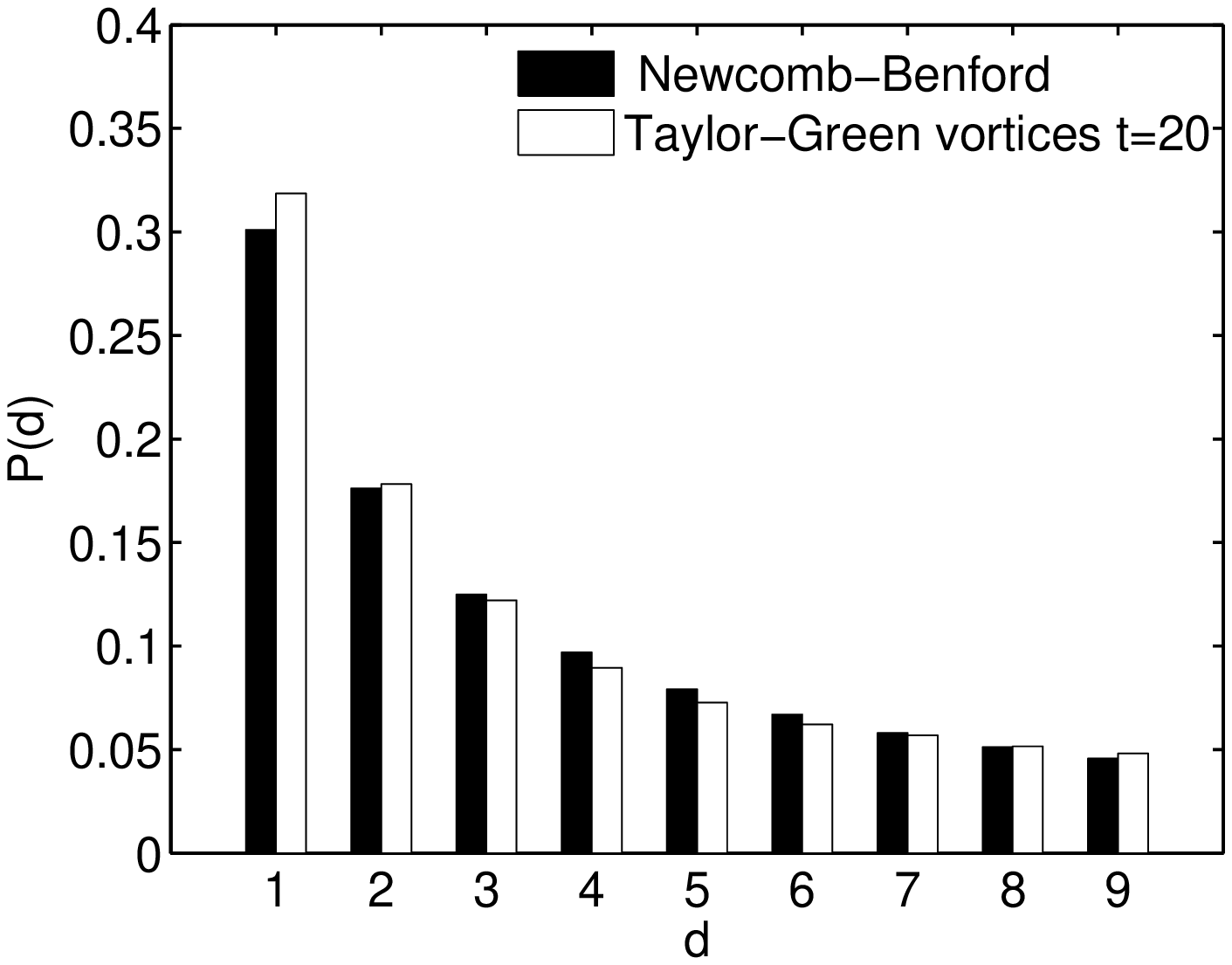}%
 \caption{\label{TGV_Histo}
Comparisons between Newcomb-Benford's law (filled bar) and the probability distribution of the first significant digits 
of dissipation for the Taylor-Green vortex (empty bar), at $t=10$ (left) and $t=20$ (right).}
\end{figure}

The transient increase of the dissipation, until its maximum value reached at $t\approx 9$, characterizes the  nonlinear vortex stretching followed by an overall decay of the flow by viscous effect. For further details see Brachet \emph{et al.} \cite{Brachet}.
Shannon's entropy slightly oscillates around the value expected from Newcomb-Benford's law, namely $H=0.8657$.
The first-digit frequencies are close to Newcomb-Benford's distribution, which is confirmed by a direct comparison shown in figure \ref{TGV_Histo}.

%-----------------------------------------------------------
\subsection{Non-homogeneous case: the plane Poiseuille flow}

Next consider a turbulent plane Poiseuille flow, namely a channel flow driven by a streamwise pressure gradient, at two different Reynolds number: 
$Re=2800$ and $Re=6880$. The Reynolds number is defined with the half-height of the channel and the bulk velocity.
The no-slip boundary condition on the walls induces a different scaling based on the friction velocity $u_\tau$ and kinematic viscosity $\nu$. Hence the wall coordinate can be expressed in wall unit: $y^+=y~\nu/u_\tau$. The two Reynolds number values based on the skin friction velocities become $Re_\tau=180$ and $Re_\tau=395$, respectively. 
The scale separation between the inner layer and the outer layer increases with Reynolds number, thus it appears useful to analyse the channel flow for two Reynolds number values. The numerical parameters are those used in Kim \emph{et al.} \cite{KMM} and Moser \emph{et al.} \cite{Moser}. 

\begin{figure}[!ht]
 \includegraphics[width=0.5\textwidth]{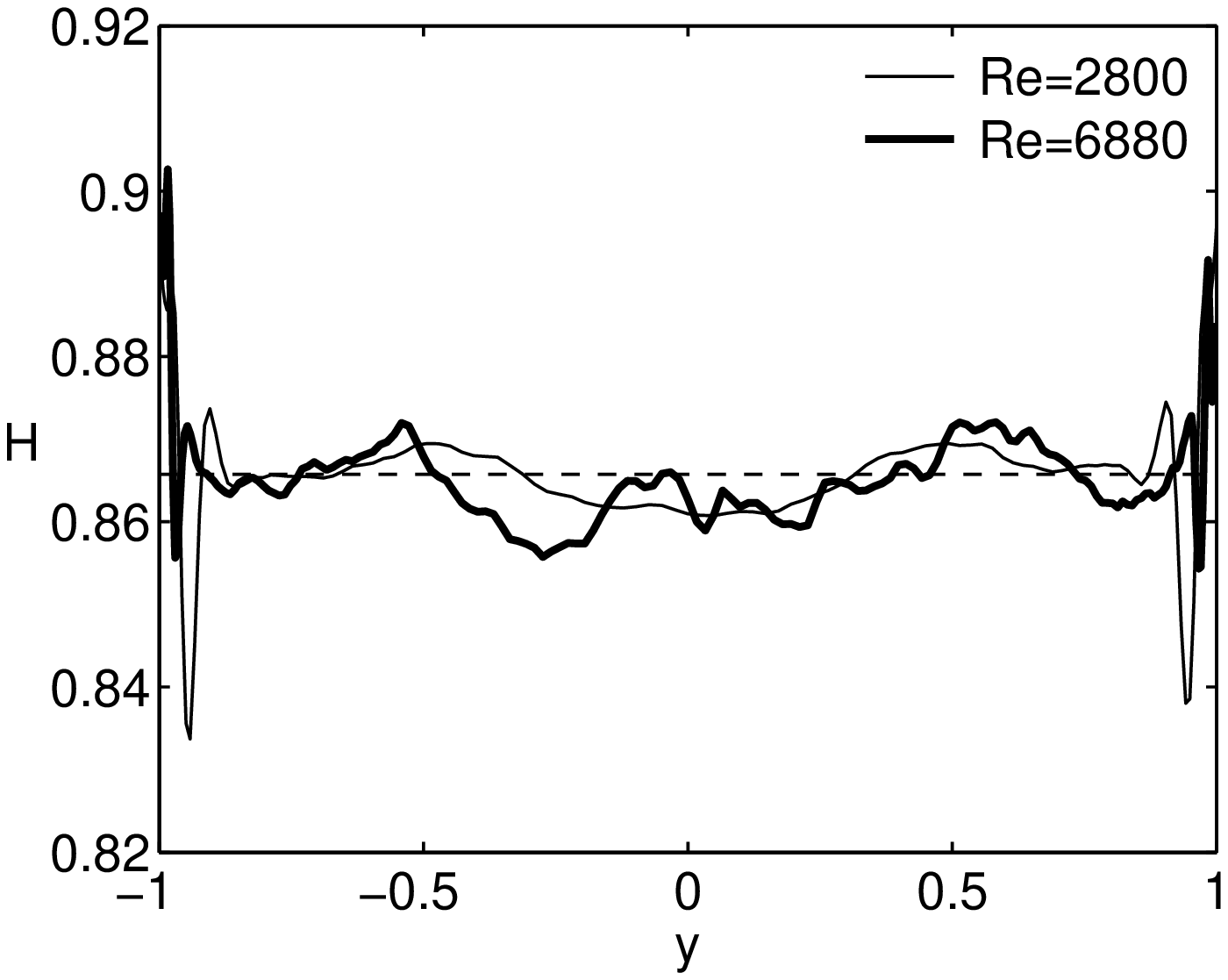}%
 \includegraphics[width=0.5\textwidth]{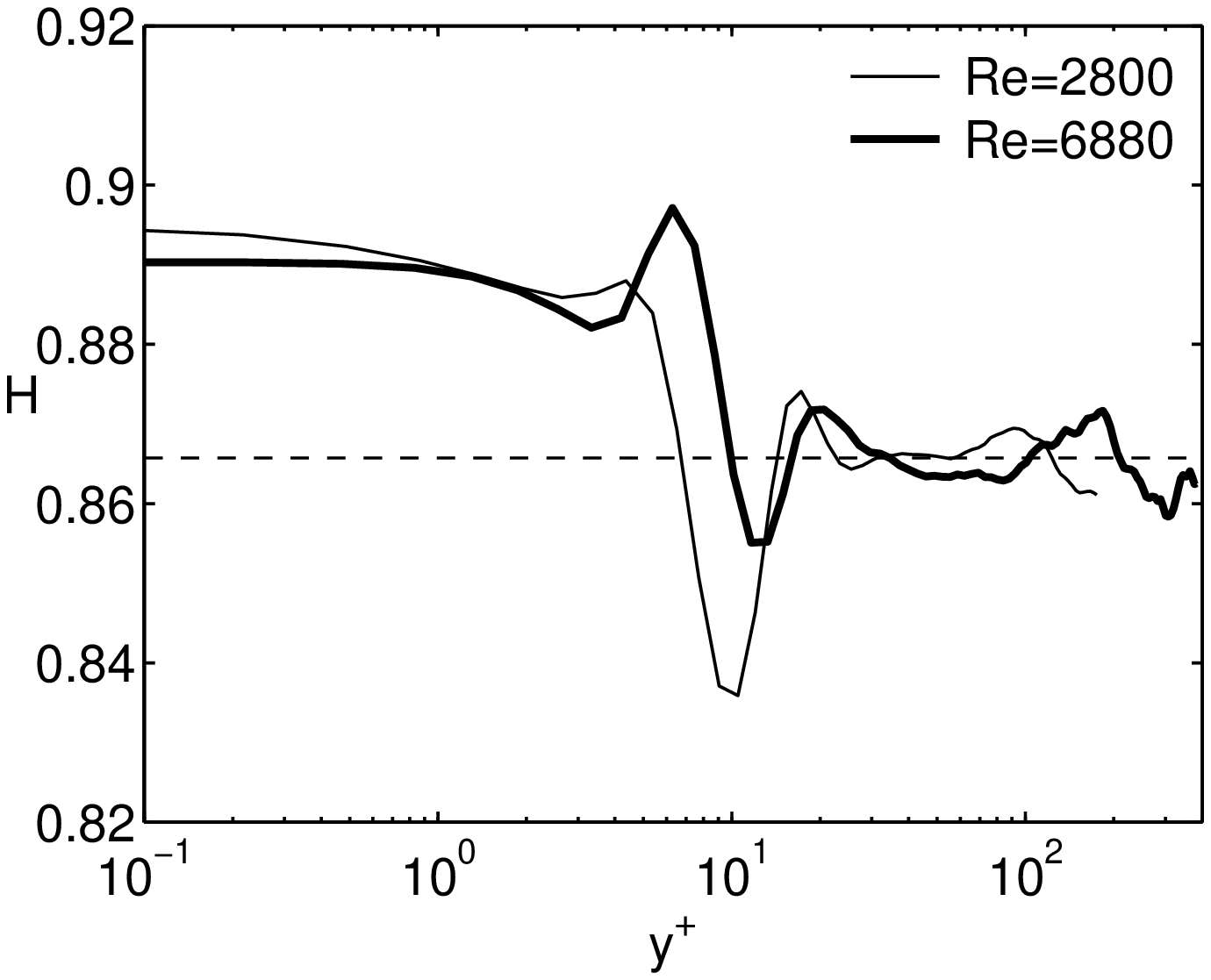}%
 \caption{\label{PPF_H_eps} 
Shannon entropy as a function of wall distance $y$ (left) and wall distance in wall units $y^+$ (right).
Results for the turbulent plane Poiseuille flow, for two Reynolds number values: Re=2800 (thin line) and Re=6880 (thick line).
The dashed line represents the entropy value corresponding to Newcomb-Benford's law, namely $H=0.8657$.}
\end{figure}

Results are temporally averaged and the conformance of turbulent statistics with Benford's law is represented, 
respectively to the wall distance expressed in outer scaling $y$ and in wall units $y^+$, on figure \ref{PPF_H_eps}.

Again, first-digit frequencies slightly oscillate around Newcomb-Benford's distribution, except in the near-wall region where the oscillation amplitudes are stronger. 
In fact, Toschi \emph{et al.} \cite{1999_PRL_Toschi} have observed maximum intermittency effects in that near-wall region, where it is known that the bursting phenomenon is the dominant dynamical feature. In addition, Toschi \emph{et al} \cite{1999_PRL_Toschi} found that for positions close to the channel centreline, scaling exponents are in substantial agreement with observations in homogeneous and isotropic turbulence. 

Shannon's entropy has been used as diagnostic tool to quantify the discrepancy with Newcomb-Benford's law.
Moreover, this information entropy can also be seen as a numerical measure which describes how informative a particular probability distribution is, ranging from zero (completely uninformative) to $H_{max}=\log_{10}~9\approx 0.9542$ (completely informative). 
With this point of view, the conclusion above is consistent with results presented by Cerbus and Goldburg \cite{Cerbus,PRE_2015_Cerbus}.
Cerbus and Goldburg observed, from experiments in a turbulent soap film, that turbulence is easier to predict where a cascade exists, or equivalently, for flows with a significant inertial range. They also found that the spatial information-entropy density is a decreasing function of the Reynolds number.

In order to  complete the previous qualitative comparisons, a quantitative comparison is presented in the table \ref{NumValues}:
\begin{table}[!ht]
\caption{\label{NumValues} Probability distribution $P(d)$ for the first significant digits. 
Distribution given by Newcomb-Benford's law (N-B) and distributions extracted from dissipation data for the 
turbulent plane Poiseuille flow.}
\begin{center}
\begin{tabular}{|c|c|c|c|c|}
\hline
digit     & Newcomb-Benford's  &  plane Poiseuille flow  &  plane Poiseuille flow \\
          &       law          &    $Re=2800$        & $Re=6880$ \\
\hline
1 &  0.3010   &  0.2892  &  0.2995 \\
2 &  0.1761   &  0.1777  &  0.1724 \\
3 &  0.1249   &  0.1297  &  0.1238 \\
4 &  0.0969   &  0.1011  &  0.0973 \\
5 &  0.0792   &  0.0820  &  0.0804 \\
6 &  0.0669   &  0.0682  &  0.0684 \\
7 &  0.0580   &  0.0579  &  0.0594 \\
8 &  0.0512   &  0.0501  &  0.0523 \\
9 &  0.0458   &  0.0442  &  0.0466 \\
\hline 
\end{tabular}
\end{center}
\end{table}

Various tests on the numerical parameters have been realized in order to check the robustness of the presented results.  
Results remain qualitatively unchanged after: varying the spatial or temporal discretization, or interpolating data on different meshes, or increasing the domain size. 
Thus presented results seem to be free from possible numerical biases. 

%%--------------------------------------------------------------------------------------------------
\section{Discussion}
%%--------------------------------------------------------------------------------------------------
%% INFORMATION
In conclusion first-digit statistics closely follow Newcomb-Benford's law. 
This is especially true when an equilibrium is reached: during the decay of an homogeneous turbulent flows or in the far from walls region of channel flows.
The scale-invariant property constituting Newcomb-Benford's law seems to be satisfied in conjunction with the eddies cascade in the inertial range.

%% application
The law of first digits is not just a mathematical curiosity, the most practical use for Newcomb-Benford's law is in fraud detection \cite{Nigrini}, 
or the possible detection of earthquakes \cite{Sambridge}.  In the framework of turbulent flows,  a practical application of first-digits statistics could be in the reduced-order models assessment, which could be investigated through their ability to conserve the first-digit frequencies (see Tolle \emph{et al.} \cite{NB_chaos}).

\begin{figure}[!ht]
\begin{center}
\includegraphics[width=0.6\textwidth]{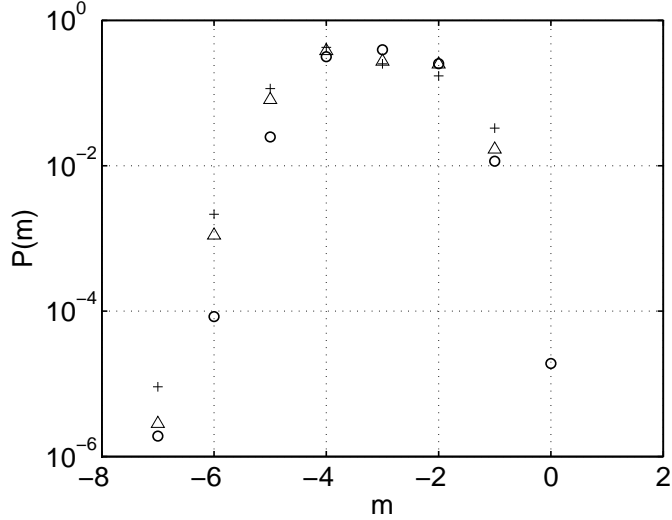}
\end{center}
\caption{\label{ProbaM} Probability mass function for the exponent $m$ in dissipation data for the 
Taylor-Green vortex at t=10 ($\circ$), Plane Poiseuille at Re=2800 ($\triangle$) and Re=6880 ($+$).}
\end{figure}

%%  P(m)
Finally the approach described in the present article can be generalized by considering the general form of scientific notation: 
\begin{equation}\label{SciNumb}
\pm~ X~ 10^m
\end{equation}
\noindent The first significant digit $d(X)$ is the object of the present study and the two other parameters (namely the sign and the exponent) can also be analysed statistically.
For the dissipation, the sign is always positive (see Equation \ref{diss}), nonetheless for other quantities, for example the velocity fluctuations, the sign-change statistics provide interesting information. Indeed the time scale related to the fluctuating velocity zero-crossing is approximately equal to the Taylor scale in turbulent flows 
(see Kailasnath \emph{et al.} \cite{1993_PoF_Kailasnath} and references within). 
The last parameter in the formula \ref{SciNumb} is the power $m$. 
The distributions of the power $m$, for the different flows considered here, are shown in the figure \ref{ProbaM}.

\noindent Despite variety of the flow configurations, the curves look very similar. However in lack of theoretical background (as far as I know), it is not possible to go deeper in the analysis, letting that as an open issue.

%--------------------------------------------------------------------------------------------------
\section*{Acknowledgement}
The author wishes to thank Eric Lamballais for the stimulating discussions on the subject and for his insightful remarks about the present manuscript.
%--------------------------------------------------------------------------------------------------

\appendix

\section{MatLab-Octave program to compute the first-digit frequencies}

function P=NewcombBenford(X); \\ 
X=abs(X);  \\ 
fd = floor(X./(10.\texttt{\^}floor(log10(X)))); \% extract the first digit \\ 
P = histc(fd,1:9)'/length(X);         \% probability

\section{MatLab-Octave program for the Taylor-Green vortex simulation}
We here consider the flow of an incompressible fluid under periodic boundary conditions with period $2\pi$. 
The pressure term can be eliminated by the incompressibility condition, then Navier-Stokes equations can be written as:
$$ \frac{\partial \textbf{u}}{\partial t} =  \textbf{u}  \times \omega - \nabla \left(p + \frac{1}{2}u^2 \right) + Re^{-1} \nabla^2 \textbf{u} $$
\noindent where $\omega=\nabla \times u$ is the vorticity.
Time marching is realized with Fourth-order Runge-Kutta method and alias error is removed by mode truncation (Orszag 2/3 rule).

\noindent {\%}{\%------------------------------------------------------------------- }\\ 
{\%}{\%		simulation of Taylor-Green Vortex }\\ 
{\%}{\%------------------------------------------------------------------- }\\ 
clear all, \\ 
Re = 1600;  \\ 
N = 2\texttt{\^}8; \\ 
dt = 5e-3; \\ 
Tend=20; \\ 
{\%------------------------------------------------------------------- }\\ 
 \\ 
{\%}{\% Fourier pseudo-spectral method }\\ 
x = 2*pi/N*[0:N-1]'; kx = [0:N/2-1 0 -N/2+1:-1]'; \\ 
for m=1:N, for n=1:N, \\
IKX(:,m,n)=i*kx; IKY(m,:,n)=i*kx; IKZ(m,n,:)=i*kx; \\
end, end\\ 
K2=-(IKX.\texttt{\^}2+IKY.\texttt{\^}2+IKZ.\texttt{\^}2); \\ 
K2p=K2; K2p(1,1,1)=1; K2p(N/2+1,:,:)=1; K2p(:,N/2+1,:)=1; K2p(:,:,N/2+1)=1;  \\ 
Z=ones(N,N,N); \\
Z = 1 - (sqrt(K2) $>$ round(N/3)+1); \\ 
Z(N/2+1,:,:)=0; Z(:,N/2+1,:)=0; Z(:,:,N/2+1)=0; \\ 
E = exp(-dt*K2/Re); E2 = exp(-dt/2*K2/Re);\\
 \\ 
{\%}{\% Newcomb-Benford's law}\\ 
NBL=log10(1+1./[1:9]); \\ 
SNB(1)=-sum(NBL.*log10(NBL)); \\ 
 \\ 
{\%}{\%  initial condition }\\ 
$\left[\mathrm{Xs,Ys,Zs}\right]$ = meshgrid(x,x,x); {\%}{\% 3D grid }\\ 
uf = fftn( 2/sqrt(3)*sin( 2*pi/3)*sin(Xs).*cos(Ys).*cos(Zs) ); \\ 
vf = fftn( 2/sqrt(3)*sin(-2*pi/3)*cos(Xs).*sin(Ys).*cos(Zs) ); \\ 
wf = zeros(N,N,N); \\ 
clear Xs Ys Zs \\ 
 \\ 
{for} k=1:round(Tend/dt) {\%}{\%===========================}\\ 
 \\ 
u=real(ifftn(uf)); v=real(ifftn(vf)); w=real(ifftn(wf)); \\ 
nltu=Z.*fftn(v.*real(ifftn(IKX.*vf-IKY.*uf))-w.*real(ifftn(IKZ.*uf-IKX.*wf))); \\ 
nltv=Z.*fftn(w.*real(ifftn(IKY.*wf-IKZ.*vf))-u.*real(ifftn(IKX.*vf-IKY.*uf))); \\ 
nltw=Z.*fftn(u.*real(ifftn(IKZ.*uf-IKX.*wf))-v.*real(ifftn(IKY.*wf-IKZ.*vf))); \\ 
pf=-(IKX.*nltu+IKY.*nltv+IKZ.*nltw)./K2p; pf(1,1,1)=0; \\ 
nltu=nltu-IKX.*pf;  ufa = E2.*(uf + dt/2*nltu); \\ 
nltv=nltv-IKY.*pf;  vfa = E2.*(vf + dt/2*nltv); \\ 
nltw=nltw-IKZ.*pf;  wfa = E2.*(wf + dt/2*nltw); \\ 
 \\ 
u=real(ifftn(ufa)); v=real(ifftn(vfa)); w=real(ifftn(wfa)); \\ 
nltua=Z.*fftn(v.*real(ifftn(IKX.*vfa-IKY.*ufa))-w.*real(ifftn(IKZ.*ufa-IKX.*wfa))); \\ 
nltva=Z.*fftn(w.*real(ifftn(IKY.*wfa-IKZ.*vfa))-u.*real(ifftn(IKX.*vfa-IKY.*ufa))); \\ 
nltwa=Z.*fftn(u.*real(ifftn(IKZ.*ufa-IKX.*wfa))-v.*real(ifftn(IKY.*wfa-IKZ.*vfa))); \\ 
pf=-(IKX.*nltua+IKY.*nltva+IKZ.*nltwa)./K2p; pf(1,1,1)=0; \\ 
nltua=nltua-IKX.*pf;  ufb = E2.*(uf + dt/2*nltua); \\ 
nltva=nltva-IKY.*pf;  vfb = E2.*(vf + dt/2*nltva); \\ 
nltwa=nltwa-IKZ.*pf;  wfb = E2.*(wf + dt/2*nltwa); \\ 
 \\ 
u=real(ifftn(ufb)); v=real(ifftn(vfb)); w=real(ifftn(wfb)); \\ 
nltub=Z.*fftn(v.*real(ifftn(IKX.*vfb-IKY.*ufb))-w.*real(ifftn(IKZ.*ufb-IKX.*wfb)));\\ 
nltvb=Z.*fftn(w.*real(ifftn(IKY.*wfb-IKZ.*vfb))-u.*real(ifftn(IKX.*vfb-IKY.*ufb)));\\ 
nltwb=Z.*fftn(u.*real(ifftn(IKZ.*ufb-IKX.*wfb))-v.*real(ifftn(IKY.*wfb-IKZ.*vfb)));\\ 
pf=-(IKX.*nltub+IKY.*nltvb+IKZ.*nltwb)./K2p; pf(1,1,1)=0; \\ 
nltub=nltub-IKX.*pf;  ufb = E.*(ufa + dt*nltub); \\ 
nltvb=nltvb-IKY.*pf;  vfb = E.*(vfa + dt*nltvb); \\ 
nltwb=nltwb-IKZ.*pf;  wfb = E.*(wfa + dt*nltwb);  \\ 
 \\ 
u=real(ifftn(ufb)); v=real(ifftn(vfb)); w=real(ifftn(wfb)); \\ 
nltuc=Z.*fftn(v.*real(ifftn(IKX.*vfb-IKY.*ufb))-w.*real(ifftn(IKZ.*ufb-IKX.*wfb))); \\ 
nltvc=Z.*fftn(w.*real(ifftn(IKY.*wfb-IKZ.*vfb))-u.*real(ifftn(IKX.*vfb-IKY.*ufb))); \\ 
nltwc=Z.*fftn(u.*real(ifftn(IKZ.*ufb-IKX.*wfb))-v.*real(ifftn(IKY.*wfb-IKZ.*vfb))); \\ 
pf=-(IKX.*nltuc+IKY.*nltvc+IKZ.*nltwc)./K2p; pf(1,1,1)=0; \\ 
nltuc=nltuc-IKX.*pf; \\ 
nltvc=nltvc-IKY.*pf; \\ 
nltwc=nltwc-IKZ.*pf; \\ 
 \\ 
uf = E.*(uf + dt/6*(nltu + 2*(nltua+nltub) + nltuc)); \\ 
vf = E.*(vf + dt/6*(nltv + 2*(nltva+nltvb) + nltvc)); \\ 
wf = E.*(wf + dt/6*(nltw + 2*(nltwa+nltwb) + nltwc)); \\ 
 \\ 
 \\ 
{\%}{\%  plot result }\\ 
time(k) = k*dt; \\ 
EPS=(2*(real(ifftn(IKX.*uf)).\texttt{\^}2+real(ifftn(IKY.*vf)).\texttt{\^}2+ ... \\
real(ifftn(IKZ.*wf)).\texttt{\^}2)+real(ifftn(IKY.*uf + IKX.*vf)).\texttt{\^}2+ ... \\
real(ifftn(IKZ.*vf + IKY.*wf)).\texttt{\^}2+real(ifftn(IKX.*wf + IKZ.*uf)).\texttt{\^}2)/Re; \\ 
Diss(k) =  mean(mean(mean(EPS)));  \\ 
X = reshape(EPS,N\texttt{\^}3,1); \\ 
fd = floor(X./(10.\texttt{\^}floor(log10(X)))); {\% extract the first digit }\\ 
stat(:,k) = histc(fd,1:9)'/length(X); {\% compute the probability }\\ 
H(k)=-sum(stat(:,k).*log10(stat(:,k))); \\ 
div = max(max(max(abs( real(ifftn(IKX.*uf+IKY.*vf+IKZ.*wf)) )))); \\ 
disp([{' time='},num2str(time(k)),{'  divergence='},num2str(div),{'  dissipation='},num2str(Diss(k))]) \\ 
subplot(1,2,1), plot(time,Diss), xlabel t, ylabel dissipation \\ 
subplot(1,2,2), plot(time,H,[0 time(k)],SNB*[1 1],{'k'}), xlabel t, ylabel H\\ 
drawnow,  \\ 
 \\ 
{end}  {\% TIME STEPPING  =================================}\\ 

\newpage
\providecommand{\noopsort}[1]{}\providecommand{\singleletter}[1]{#1}%


\begin{thebibliography}{10}
\expandafter\ifx\csname url\endcsname\relax
  \def\url#1{\texttt{#1}}\fi
\expandafter\ifx\csname urlprefix\endcsname\relax\def\urlprefix{URL }\fi
\expandafter\ifx\csname href\endcsname\relax
  \def\href#1#2{#2} \def\path#1{#1}\fi

\bibitem{Newcomb}
S.~Newcomb, Note on the frequency of use of the different digits in natural
  numbers, Am. J. Math. 4 (1881) 39--40.

\bibitem{Benford}
F.~Benford, The law of anomalous numbers, Proc. Am. Philos. Soc. 78 (1938)
  551--572.

\bibitem{Hill98}
T.~P. Hill, The first digit phenomenon, Am. Sci. 86 (1998) 358--363.

\bibitem{Sambridge}
M.~Sambridge, H.~Tkalcic, A.~Jackson, Benford’s law in the natural sciences,
  Geophys. Res. Lett. 37 (2010) L22301.

\bibitem{Raimi}
R.~A. Raimi, The first digit problem, Am. Math. Monthly 83 (1976) 521--538.

\bibitem{Hill95}
T.~P. Hill, A statistical derivation of the significant-digit law, Statist.
  Sci. 10 (1995) 354--363.

\bibitem{Pietronero}
L.~Pietronero, E.~Tosatti, V.~Tosatti, Vespignani, Explaining the uneven
  distribution of numbers in nature: the laws of {B}enford and {Z}ipf, Physica
  A 293 (2001) 297--304.

\bibitem{Pinkham}
R.~S. Pinkham, On the distribution of first significant digits, Ann. Math.
  Statist. 32 (1961) 1223--1230.

\bibitem{Pocheau}
A.~Pocheau, The significant digit law: a paradigm of statistical scale
  symmetries, Eur. Phys. J. B 49 (2006) 491--511.

\bibitem{Fewster}
R. M.~Fewster A Simple Explanation of Benford's Law, 
The American Statistician 63 (2009) 26-32.


\bibitem{Luque}
B.~Luque, L.~Lacasa, The first-digit frequencies of prime numbers and {R}iemann
  zeta zeros, Proc. R. Soc. A 465 (2009) 2197--2216.

\bibitem{Frisch}
U.~Frisch, Turbulence. The legacy of A. N. Kolmogorov., Cambridge University
  Press, 1995.

\bibitem{Vassilicos}
J.~C. Vassilicos, Dissipation in turbulent flows, Annual Review of Fluid
  Mechanics 47 (2015) 95--114.

\bibitem{Shannon}
C.~E. Shannon, A mathematical theory of communication, Bell Syst. Tech. J. 27
  (1948) 379--423.

\bibitem{TGV}
G.~I. Taylor, A.~E. Green, Mechanism of the production of small eddies from
  large ones, Proc. Roy. Soc. A, Mathematical and Physical Sciences, 158 (1936)
  499--521.

\bibitem{Brachet}
M.~E. Brachet, D.~I. Meiron, S.~A. Orszag, B.~G. Nickel, R.~H. Morf, U.~Frisch,
  Small-scale structure of the Taylor-Green vortex, J. Fluid Mech. 130 (1983) 411--452.

\bibitem{KMM}
J.~Kim, P.~Moin, R.~Moser, Turbulence statistics in fully developed channel
  flow at low reynolds number, J. Fluid Mech. 177 (1987) 133--166.

\bibitem{Moser}
R.~D. Moser, J.~Kim, N.~N. Mansour, Direct numerical simulation of turbulent
  channel flow up to {R}e= 590, Phys. Fluids 11 (1999) 943--945.

\bibitem{1999_PRL_Toschi}
F.~Toschi, G.~Amati, S.~Succi, R.~Benzi, R.~Piva, Intermittency and structure
  functions in channel flow turbulence, Phys. Rev. Lett 82 (1999) 5044.

\bibitem{Cerbus}
R.~T. Cerbus, W.~I. Goldburg, Information content of turbulence, Phys. Rev. E
  88 (2013) 053012.

\bibitem{PRE_2015_Cerbus}
R.~T. Cerbus, W.~I. Goldburg, Predicting two-dimensional turbulence, Phys. Rev.
  E 91 (2015) 043003.

\bibitem{Nigrini}
M.~Nigrini, A taxpayer compliance application of {B}enford's law, J. Amer. Tax
  Assoc. 18 (1996) 72--91.

\bibitem{NB_chaos}
C.~R. Tolle, J.~L. Budzien, R.~A. LaViolette, Do dynamical systems follow
  {B}enford's law?, Chaos 10 (2000) 331--336.

\bibitem{1993_PoF_Kailasnath}
P.~Kailasnath, K.~R. Sreenivasan, Zero crossings of velocity fluctuations in
  turbulent boundary layers, Phys. Fluids 5 (1993) 2879.

\end{thebibliography}
\end{document}